\documentstyle[multicol,aps,epsf]{revtex}

\begin{document}
\title{A Computational Fluid Model for Investigation of Plasma Waves and
Instabilities }
\author{H. Hakimi Pajouh$^1$\footnote{hakimi@theory.ipm.ac.ir}, M. R. Rouhani$^2$, H. Abbasi$^{1,3}$,
F. Kazeminejad$^4$, and S. Rouhani$^5$}
\address{$^1$Institute for Studies in Theoretical Physics and Mathematics,
P. O. Box 19395-5531, Tehran, Iran \\
$^2$Depatment of Physics, Az-zahra University, P. O. Box 19834, Tehran, Iran\\
$^3$Faculty of Physics, Amir Kabir University of Technology, P. O. Box 15875-4413, Tehran, Iran\\
$^4$Independent consultant\\
$^5$Faculty of Physics, Sharif University of Technology, P. O. Box 11365-9161, Tehran, Iran}
\maketitle
\begin{abstract}
A computational fluid model is developed to study waves and
instabilities. A new technique involving initial perturbations in
configuration space have been implemented to excite the plasma
waves; i.e. the perturbations acting similar to a random velocity
distribution in particle in cell (PIC) codes. This forms a new
powerful tool for investigation of many waves arising in both
thermal and cold plasmas and as such will allow investigation of
problems demanding scales and resolution not yet possible by PIC
codes. The model predicts Langmuir waves, two stream
instabilities, nonlinear wave-wave interaction, and the Debye
screening effects. The agreement between theory and simulation
where analytic results are available are excellent.
\end{abstract}

\section{INTRODUCTION}

Due to the fundamental importance of the waves and instabilities
in plasma and hydrodynamics investigations, computational
researchers have devoted great efforts in developing appropriate
tools. One of the main challenges after developing numerically
stable algorithms in fluid models has been generation of the waves
in the linear, nonlinear as well as unstable modes; i.e. waves
which preserve analytic dispersion relations\footnote{The
waves'propagation characteristics are encoded in the dispersion
relations\cite{whitham}}\cite{Tam}. Furthermore extending the case
of hydrodynamics to that of MHD and or plasma physics one deals
with waves with considerably more complicated propagation
characteristics than the hydrodynamics cases treated by those
authors; i.e. dispersion, polarization, oblique propagations, etc.

The main problems in generating a wave spectrum from small
amplitude disturbances in fluid equations are: (1) the highly
nonlinear nature of those equations; (2) the lack of an initial
thermal velocity distribution. The first problem could cause any
small amplitude configuration space disturbance to grow to very
large amplitudes in relatively short times and result in wave
breaking and non-propagation. Also when there does exist a thermal
distribution, there are always a distribution of thermalized
particles in phase with most waves; they can therefore excite the
allowed modes to at least half their thermal level. Therefore in a
case without thermal equilibrium, a disturbance of arbitrary
wavelength cannot strictly speaking apportion its energy to other
allowed modes. For example in purely electrostatic cases, we know
from equilibrium statistical mechanics that when there exist  a
thermal distribution each mode $E_l(k)$ can acquire an energy
\cite{dawson}:
\begin{equation}
\frac{<\mid E_l(k)\mid^2>}{8\pi}\propto kT.
\end{equation}
To investigate MHD wave spectra therefore magnetohydrodynamic
particle codes have served as powerful tools\cite{lebof},
\cite{tajima}, \cite{brunel}, and \cite{kazemi}. For other plasma
waves PIC \cite{birdsal} and \cite{hockney} or hybrid codes \cite
{kazemi}, \cite{winske}, and \cite{hono} have served as the main
wave investigation tools; i.e., basically codes which start from
thermal equilibrium. In these codes the  random particle
distribution acts like a disturbance in velocity space and
configuration space remains unaltered at the beginning of each
simulation.

In our case we initiate each simulation by a perturbation in
configuration space. Despite the initial shape of the
perturbation, we observe other allowed modes to develop similar to
PIC simulations. We believe that the mesh discretization and the
finite differencing contribute in the following ways: (i) round of
errors alter the initial perturbation shape and can drive other
wavelength; (ii) as the nonlinear effects grow amplitudes and
shorten wavelengths to the numerical dissipation and dispersion
scale lengths, these effects can act to dampen and initiate the
propagation of the different modes and prevent indefinite
nonlinear growth. These effects can therefore explain the observed
wave spectra. With this then we can use fluid instead of PIC codes
as a convenient alternative to investigate many waves.

The organization of the paper is as follows: in section II the
model is treated analytically; in section III the numerical scheme
(algorithm, stability and conservation laws) are presented; in
section IV the various tests of the model are presented (test of
the dispersion relation, two stream instability, screening effect
and nonlinear harmonic generation). At the end a brief summary and
conclusion with future direction are presented.

\section{ANALYTICAL\ TREATMENT}

We focus on the investigation of the high frequency (hf)
longitudinal waves; i. e. a frequency domain where ions can be
safely assumed to form an immobile background ($n_{0}$ represents
their uniform density). The appropriate equations are then
Poisson's and the electron fluid equations:
\begin{equation}
\frac{\partial n}{\partial t}+\frac{\partial }{\partial x}(nv)=0,  \label{1}
\end{equation}
\begin{equation}
\frac{\partial v}{\partial t}+v\frac{\partial }{\partial x}v=\frac{e}{m}%
\frac{\partial }{\partial x}\varphi -\frac{1}{nm}\frac{\partial P}{\partial x%
},  \label{2}
\end{equation}
\begin{equation}
\frac{\partial ^{2}\varphi }{\partial x^{2}}=4\pi e(n-n_{0}).  \label{3}
\end{equation}
Here $\varphi $ is the self-consistent electric potential, and
$n$, $v$, $P$ and $m$ represent the electron density, velocity,
pressure and rest mass respectively. Without any loss of
generality this problem is treated in one dimension. These basic
equations are supplemented by an ''equation of state'' according
to the particular thermodynamic properties of the fluid of
interest. Here, isothermal equation of state is used:
\begin{equation}
P=nT,  \label{4}
\end{equation}
where $T$ is the electron temperature and is assumed to be
constant and Boltzmann's constant, $k$, is assumed to be unity.

The minimum requirement of any computational model lies in its
ability to preserve conservation laws; for that fluid equations
are cast in flux conservative form. Equation (\ref{2}) in
conservative form upon using Eq. (\ref{4}) in Eq. (\ref{2})
becomes:
\begin{equation}
\frac{\partial v}{\partial t}+\frac{\partial }{\partial x}\left( \frac{1}{2}%
v^{2}-\frac{e}{m}\varphi +\frac{T}{m}\ln n\right) =0.  \label{5}
\end{equation}
Note that the logarithmic term is caused by the electron pressure.
Therefore the three equations that form the basis of our model
are:
\begin{equation}
\frac{\partial n}{\partial t}+\frac{\partial }{\partial x}(nv)=0,  \label{6}
\end{equation}
\begin{equation}
\frac{\partial v}{\partial t}+\frac{\partial }{\partial x}\left( \frac{1}{2}%
v^{2}-\frac{e}{m}\varphi +\frac{T}{m}\ln n\right) =0,  \label{7}
\end{equation}
\begin{equation}
\frac{\partial ^{2}\varphi }{\partial x^{2}}=4\pi e(n-n_{0}).  \label{8}
\end{equation}
We will next derive a dispersion relation for wave propagation using Eqs. (%
\ref{6}), (\ref{7}), and (\ref{8}). To do this, linearizing Eqs.
(\ref{6}), (\ref{7}), and (\ref {8}) about a spatially uniform
equilibrium ($n=n_{0}+\delta n$, $v=\delta v$ and $\varphi =\delta
\varphi $), we obtain the following set:
\begin{equation}
\frac{\partial \delta n}{\partial t}+n_{0}\frac{\partial }{\partial x}\delta
v=0,  \label{9}
\end{equation}
\begin{equation}
\frac{\partial \delta v}{\partial t}+\frac{\partial }{\partial x}\left( -%
\frac{e}{m}\delta \varphi +\frac{1}{n_{0}}\delta n\right) =0,  \label{10}
\end{equation}
\begin{equation}
\frac{\partial ^{2}\delta \varphi }{\partial x^{2}}=4\pi e\delta n.
\label{11}
\end{equation}
Assuming simple plane wave solutions, Eqs. (\ref{9}), (\ref{10}),
and (\ref{11}) reduce to the following set of equations:
\begin{equation}
-i\omega \delta n+ikn_{0}\delta v=0,  \label{12}
\end{equation}
\begin{equation}
-i\omega \delta v+ik(-\frac{e}{m}\delta \varphi +\frac{1}{n_{0}}\delta n)=0,
\label{13}
\end{equation}
\begin{equation}
-k^{2}\delta \varphi =4\pi e\delta n.  \label{14}
\end{equation}
Eqs. (\ref{12}), (\ref{13}), and (\ref{14}) yield nontrivial
solution if the following is obeyed:
\begin{equation}
\omega ^{2}=\omega _{p}^{2}+k^{2}v_{T}^{2},  \label{15}
\end{equation}
where
\begin{equation}
\omega _{p}^{2}=\frac{4\pi e^{2}n_{0}}{m}\text{ and }v_{T}^{2}=\frac{T}{m}
\label{16}
\end{equation}
are the electron plasma frequency and the thermal velocity,
respectively.

Studies of Langmuir waves (hf electron waves) are of particular importance.
Aside from the applications to real experimental situations which will
become evident in the application section, they serve as excellent probes
for testing the validity of the fluid code that we have developed.

\section{NUMERICAL\ ALGORITHM}

Our model is simply an intuitive construct based on well-known
fluid dynamics and Poisson's equations, geared toward plasma
physics applications, where many different wave phenomena in
dispersive media are of interest. Its physical ''conceptual
basis'' can be regarded as a model that treats non-stationary
electron wave motion for hf domain where $\omega \gg kv_{T}$ in
linear and nonlinear regions. Besides, it can predict electron
wave spectrum more accurately than ''particle in cell simulation''
as here we expect less numerical noise.

\subsection{Normalization}

In these calculations we use the following normalizations:
\begin{equation}
\omega _{p}t\rightarrow t,\quad \frac{x}{r_{D}}\rightarrow x,\quad \frac{v}{%
v_{T}}\rightarrow v,\quad \frac{n}{n_{0}}\rightarrow n,\quad \frac{e\varphi
}{T}\rightarrow \varphi ,\quad  \label{17}
\end{equation}
where
\begin{equation}
\text{ }r_{D}^{2}=\frac{T}{4\pi e^{2}n_{0}}  \label{18}
\end{equation}
is the electron Debye length. Using these definitions, Eqs. (\ref{1}), (\ref{3}), (\ref{5}%
), and (\ref{15}) can now be rewritten as follows:
\begin{equation}
\frac{\partial n}{\partial t}+\frac{\partial }{\partial x}(nv)=0,  \label{19}
\end{equation}
\begin{equation}
\frac{\partial v}{\partial t}+\frac{\partial }{\partial x}\left( \frac{1}{2}%
v^{2}-\varphi +\ln n\right) =0,  \label{20}
\end{equation}
\begin{equation}
\frac{\partial ^{2}\varphi }{\partial x^{2}}=n-1,  \label{21}
\end{equation}
\begin{equation}
\omega ^{2}=1+k^{2}.  \label{22}
\end{equation}
It is already mentioned, logarithmic term in Eq. (\ref{20}) is caused by the
electron pressure.. Thus the code has the flexibility of being easily
converted to the case when electron pressure is negligible.

\subsection{The Numerical Scheme}

Next we shall describe the numerical scheme. The steps of the
scheme are summarized in Table I. A Lax-Wendroff method is used to
push $n$ and $v$, while a poisson solver at the end of each step
updates the electric potential.

The grid spacing and time step are denoted by $\Delta $ and $\Delta t$
respectively. The fluid velocity and density are known at integer time step $%
l$. To complete the initial conditions, $\varphi $ is computed at
the same time step ($l$) by the help of a Poisson solver that is
based on tridiagonal matrix method. Then $n$ and $v$ are pushed
from $l$ to $l+1/2$ as the auxiliary step of the Lax-Wendroff
scheme using Eqs. (\ref{19}) and (\ref{20}) (please refer to item
3 of the Table I). Then again $\varphi $ is computed in the
auxiliary step ($l+1/2$) using the value of $n$ in the mentioned
step. Having known $n$, $v$, and $\varphi $ at the time step
$l+1/2$, we push $n$ and $v$ all the way to time step $l+1$ as the
main step of the Lax-Wendroff scheme in Eqs. (\ref{19}) and
(\ref{20}) (items 5 and 6 in Table I). The electric potential
$\varphi $ is then computed at the time step $l+1$ using $n^{l+1}$.
\begin{center}
\begin{tabular}{|l|}
\hline
\begin{tabular}{l}
\quad \quad \quad \quad \quad \quad \quad \quad \quad \quad \quad \quad
TABLE\ I \\
\quad \quad \quad Numerical Algorithm of the Fluid Model for Plasma Waves
\end{tabular}
\\ \hline
\begin{tabular}{l}
Initially we have: $n_{m}^{l}$, $v_{m}^{l}$ \\
\quad 1. Compute electric potential, $\varphi _{m}^{l}$, using Poisson
solver. \\
\quad 2. Compute fluxes in continuity and momentum equation in main step: \\
\quad $\quad (f_{n})_{m}^{l}=n_{m}^{l}v_{m}^{l},$ \\
\quad \quad $(f_{v})_{m}^{l}=\frac{1}{2}(v_{m}^{l})^{2}-\varphi _{m}^{l}+\ln
n_{m}^{l}.$ \\
\quad 3. Push velocity and density half a time step: \\
\quad \quad $n_{m+1/2}^{l+1/2}=\frac{1}{2}(n_{m+1}^{l}+n_{m}^{l})-\frac{%
\Delta t}{2\Delta }\left[ (f_{n})_{m+1}^{l}-(f_{n})_{m}^{l}\right] ,$ \\
\quad $\quad v_{m+1/2}^{l+1/2}=\frac{1}{2}(v_{m+1}^{l}+v_{m}^{l})-\frac{%
\Delta t}{2\Delta }\left[ (f_{v})_{m+1}^{l}-(f_{v})_{m}^{l}\right] .$ \\
\quad 4. Compute electric potential in half step, $\varphi
_{m+1/2}^{l+1/2}$, using $n_{m+1/2}^{l+1/2}$.
\\
\quad 5. Compute fluxes in continuity and momentum equations in half step:
\\
$\quad \quad (f_{n})_{m+1/2}^{l+1/2}=n_{m+1/2}^{l+1/2}v_{m+1/2}^{l+1/2},$ \\
\quad \quad $(f_{v})_{m+1/2}^{l+1/2}=\frac{1}{2}(v_{m+1/2}^{l+1/2})^{2}-%
\varphi _{m+1/2}^{l+1/2}+\ln n_{m+1/2}^{l+1/2}.$ \\
\quad 6.Push the velocity and density another half a time step: \\
$\quad \quad n_{m}^{l+1}=n_{m}^{l}-\frac{\Delta t}{\Delta }\left[
(f_{n})_{m+1/2}^{l+1/2}-(f_{n})_{m-1/2}^{l+1/2}\right] ,$ \\
\quad $\quad v_{m}^{l+1}=v_{m}^{l}-\frac{\Delta t}{\Delta }\left[
(f_{v})_{m+1/2}^{l+1/2}-(f_{v})_{m-1/2}^{l+1/2}\right] $.\\
\quad 7.Compute electric potential in the main step, $\varphi
_{m}^{l+1}$, using $n_{m}^{l+1}$.\\
\end{tabular}
\\ \hline
\end{tabular}
\end{center}

\subsection{Conservation Laws}

Equations (19), (20) are in conservative form, and we demand that
the corresponding difference equations to be equally conservative.
More specifically, we expect finite difference scheme to conserve
the mass density ($\int_{-\infty}^{+\infty} n dx$), momentum and
the energy of the system, irrespective of the errors incurred by
the finite difference lattice.

To investigate the conservation laws, in what follows, a method
compatible with both the auxiliary and the main steps will be
presented \cite{Potter}. That is, Eqs. (\ref{19}) and (\ref{20})
are integrated over each space-time cell ($m$) of area $\Delta
t\Delta_m $ ($\Delta t=t^{l+1}-t^{l}$) as follows:
\begin{equation}
\int_{t^{l}}^{t^{l+1}}dt\int_{\Delta _{m}}dx\frac{\partial n}{\partial t}%
=-\int_{t^{l}}^{t^{l+1}}dt\int_{\Delta _{m}}dx\frac{\partial }{\partial x}%
(nv),  \label{23}
\end{equation}
\begin{equation}
\int_{t^{l}}^{t^{l+1}}dt\int_{\Delta _{m}}dx\frac{\partial v}{\partial t}%
=-\int_{t^{l}}^{t^{l+1}}dt\int_{\Delta _{m}}dx\frac{\partial }{\partial x}%
\left( \frac{1}{2}v^{2}-\varphi +\ln n\right) .  \label{24}
\end{equation}
Here $\int_{\Delta _{m\text{ }}}$denotes integral over the cell labelled by $%
m $. Carrying out trivial integration over $dt$ and $dx$ on the
left and right sides respectively Eqs. (\ref{23}) and (\ref{24})
become:
\begin{equation}
\int_{\Delta _{m}}n^{l+1}dx-\int_{\Delta
_{m}}n^{l}dx=-\int_{t^{l}}^{t^{l+1}}dt\sum_{\alpha }(nv)_{m},  \label{25}
\end{equation}
\begin{equation}
\int_{\Delta _{m}}v^{l+1}dx-\int_{\Delta
_{m}}v^{l}dx=-\int_{t^{l}}^{t^{l+1}}dt\sum_{\alpha }\left( \frac{1}{2}%
v^{2}-\varphi +\ln n\right) _{m},  \label{26}
\end{equation}
where $\alpha $ stands for the boundaries of every cell (the right
and the left). Using
\begin{equation}
\int_{\Delta _{m}}\left(
\begin{array}{l}
n^{l} \\
v^{l}
\end{array}
\right) dx =\Delta \left(
\begin{array}{l}
n_{m}^{l} \\
v_{m}^{l}
\end{array}
\right).  \label{27}
\end{equation}
the following equations are thus obtained:
\begin{equation}
n_{m}^{l+1}=n_{m}^{l}-\int_{t^{l}}^{t^{l+1}}dt\frac{1}{\Delta }\sum_{\alpha
}(nv)_{m}  \label{28}
\end{equation}
\begin{equation}
v_{m}^{l+1}=v_{m}^{l}-\int_{t^{l}}^{t^{l+1}}dt\frac{1}{\Delta
}\sum_{\alpha }\left( \frac{1}{2}v^{2}-\varphi +\ln n\right) _{m}.
\label{29}
\end{equation}
Summing over cells ($m$) in the system results in:
\begin{equation}
\sum_{m=1}^{M}\left( n_{m}^{l+1}-n_{m}^{l}\right)
=-\sum_{m=1}^{M}\int_{t^{l}}^{t^{l+1}}dt\frac{1}{\Delta }\sum_{\alpha
}(nv)_{m},  \label{30}
\end{equation}
\begin{equation}
\sum_{m=1}^{M}\left( v_{m}^{l+1}-v_{m}^{l}\right)
=\sum_{m=1}^{M}\int_{t^{l}}^{t^{l+1}}dt\frac{1}{\Delta }\sum_{\alpha }\left(
\frac{1}{2}v^{2}-\varphi +\ln n\right) _{m}.  \label{31}
\end{equation}
Since finite differences were used in computing all the
derivatives, then if one sums over all the grid cells in the
system, each such quantities will appear twice with opposite signs
corresponding to the cell boundaries that are being shared between
the neighboring cells, and they will thus add up to zero. There
can, however, be contributions from the walls of the computation
box. For the periodic boundary condition the walls contributions
gives zero; for other cases appropriate boundary conditions are
implemented to insure good conservation using guard cells.

\subsection{Numerical Stability Analysis}

In order to obtain the Courant-Fredricks-Lewy (CFL) condition for
the model, the difference equations (obtained from the
differential equations for the problem by discretizing them) must
be considered. We follow the method of Potter \cite{Potter};
\textit{i.e.} obtain the integration time pusher operator from the
difference equations assuming a spatially uniform system and solve
them in Fourier space and obtain a non-local result. We shall do
the stability analysis with the pressure term.

Recall that the differential equations (\ref{19}), (\ref{20}) and
(\ref{21}) formed the basis of the model. These equations upon
linearization, give:
\begin{equation}
\frac{\partial \delta n}{\partial t}+\frac{\partial }{\partial x}\delta v=0,
\label{32}
\end{equation}
\begin{equation}
\frac{\partial \delta v}{\partial t}+\frac{\partial }{\partial x}(-\delta
\varphi +\delta n)=0,  \label{33}
\end{equation}
\begin{equation}
\frac{\partial ^{2}\delta \varphi }{\partial x^{2}}=\delta n.  \label{34}
\end{equation}
Next using Eqs. (\ref{32}), (\ref{33}), and (\ref{34}), after
combining the
auxiliary and the main steps of the Lax-Wendroff scheme and assuming $n$, $v$%
, and $\varphi $ to have the form ($l$ refers to the time step and $m$
inside the parenthesis to the grid location along $x$)
\begin{equation}
(n^{l}\text{, }v^{l}\text{, }\varphi ^{l})=(\hat{n}^{l}\text{, }\hat{v}^{l}%
\text{, }\hat{\varphi}^{l})e^{i(km\Delta )},  \label{35}
\end{equation}
we obtain the following integration matrix ($\sigma =k\Delta /2$):
\begin{equation}
\left(
\begin{array}{l}
n \\
v \\
\varphi
\end{array}
\right) ^{l+1}=\left(
\begin{array}{ccc}
1-\frac{2\Delta t^{2}}{\Delta ^{2}}\sin ^{2}\sigma & \frac{-2i\Delta t}{%
\Delta }\sin \sigma \cos \sigma & \frac{2\Delta t^{2}}{\Delta ^{2}}\sin
^{2}\sigma \\
\frac{-2i\Delta t}{\Delta }\sin \sigma \cos \sigma -\frac{i\Delta \Delta t}{2%
}\cot \sigma & 1-\frac{\Delta t^{2}}{2}-\frac{2\Delta t^{2}}{\Delta ^{2}}%
\sin ^{2}\sigma & 0 \\
\frac{\Delta t^{2}}{2}-\frac{\Delta ^{2}}{4\sin ^{2}\sigma } & \frac{i\Delta
\Delta t}{2}\cot \sigma & -\frac{\Delta t^{2}}{2}
\end{array}
\right) \left(
\begin{array}{l}
n \\
v \\
\varphi
\end{array}
\right) ^{l}
\end{equation}
Thus, according to Von Neumann stability condition the following
inequality should be held:\footnote{$()^{l+1}=g()^l$ where
$g=e^{-i \omega \Delta t}$; Von Neumann stability condition holds
for $\omega$ real.}
\begin{equation}
\left| g_{\mu }\right| \leq 1,  \label{37}
\end{equation}
where $g_{\mu }$ are the eigenvalues of the integration matrix and
subscript refer to different eigenvalues (here $\mu =1,2,3$). The
value of $g_{\mu }$ is then determined by setting the following
determinant equal to zero; i.e.,
\begin{equation}
\left|
\begin{array}{ccc}
1-\frac{2\Delta t^{2}}{\Delta ^{2}}\sin ^{2}\sigma -g & \frac{-2i\Delta t}{%
\Delta }\sin \sigma \cos \sigma & \frac{2\Delta t^{2}}{\Delta ^{2}}\sin
^{2}\sigma \\
\frac{-2i\Delta t}{\Delta }\sin \sigma \cos \sigma -\frac{i\Delta \Delta t}{2%
}\cot \sigma & 1-\frac{\Delta t^{2}}{2}-\frac{2\Delta t^{2}}{\Delta ^{2}}%
\sin ^{2}\sigma -g & 0 \\
\frac{\Delta t^{2}}{2}-\frac{\Delta ^{2}}{4\sin ^{2}\sigma } & \frac{i\Delta
\Delta t}{2}\cot \sigma & -\frac{\Delta t^{2}}{2}-g
\end{array}
\right| =0  \label{38}
\end{equation}
The corresponding solutions for $g$ are simply:
\[
g_{1}=0,
\]
\begin{equation}
g_{2,3=}1-\frac{1}{2}\Delta t^{2}-\frac{2\Delta t^{2}}{\Delta ^{2}}\sin
^{2}\sigma \pm i\sqrt{\Delta t^{2}\cos ^{2}\sigma \left( 1+\frac{4}{\Delta
^{2}}\sin ^{2}\sigma \right) }.  \label{39}
\end{equation}
$g_{1}$ fulfills the inequality (\ref{37}). For the two other eigenvalues,
we have:
\begin{equation}
\left| g_{2}\right| =\left| g_{3}\right| =\left[ 1-\Delta t^{2}\left( 1+%
\frac{4}{\Delta ^{2}}\sin ^{2}\sigma \right) +\Delta t^{4}\left( \frac{1}{4}%
+\cos ^{4}\sigma \right) \left( 1+\frac{4}{\Delta ^{2}}\sin ^{2}\sigma
\right) ^{2}\right] ^{1/2}  \label{40}
\end{equation}
Equation (\ref{37}) is then obeyed if the following inequality is
held:
\begin{equation}
\Delta t^{2}\left( \frac{1}{4}+\cos ^{4}\sigma \right) \left( 1+\frac{4}{%
\Delta ^{2}}\sin ^{2}\sigma \right) \leq 1.  \label{41}
\end{equation}
Since $\Delta t$ and $\Delta $ are small values ($0<\Delta t\ll 1$ and $%
0<\Delta \ll 1$) the inequality (\ref{41}) will be satisfied if$:$
\begin{equation}
\frac{\Delta t}{\Delta }\leq \frac{2}{\sqrt{4+\Delta ^{2}}}.  \label{42}
\end{equation}
Inequality (\ref{42}) is exact up to the scheme accuracy, however,
taking into account the smallness of $\Delta t$ and $\Delta $
the following stability condition results:
\[
\frac{\Delta t}{\Delta }\leq 1.
\]
\section{TESTING\ THE\ CODE}

As mentioned, we have constructed the one-dimensional version of
the code and have tested it by looking at small and large
amplitude (nonlinear) effects in an initially uniform plasma. In
what follows, a review of the results will be given.

\subsection{Dispersion relation}

The most basic requirement of a computational model aside from
conservation laws is its ability to predict the linear theory;
e.g. the waves dispersion relation. The degree to which the
analytic dispersion relation is obeyed acts as a gauge of the
computational model and serves to determine its limitations.

From Eq. 40, the dispersion relation of the corresponding difference
equation is:
$$
e^{-\omega _{I}\Delta t}\sin (\omega _{R}\Delta t)=\sqrt{(\Delta t)^{2}\cos
^{2}\sigma (1+\frac{4}{\Delta ^{2}}\sin ^{2}\sigma )}.
$$
where $\omega=\omega_R + \omega_I$. Comparison of this with the
analytic dispersion relation shows that by changing
$k\longrightarrow k\sin (k\Delta )/(k\Delta )$ in the analytic
case
one roughly recovers the above result for $\Delta t\ \omega _{R}\ll 1$ , $%
k\Delta \ll 1$ . The fact that $\omega _{I}$ does not have any $k$
dependence, implies no part of the $k$ space to be more
susceptible to
numerical instability than others\footnote{%\label{foota}
Many PIC algorithms show $\omega _{I}\propto k^{2}$; i.e. intense
short wavelength noise or instability.}. The difference dispersion
relation above also indicates that for $\sin (k\Delta )/(k\Delta
)\longrightarrow 1$ the numerical dispersion to disappear; i.e.
for modes with wavelengths long compared with the grid spacing it
should be negligible.

For the initial perturbations, small fluctuations in the density from a
uniform background were implemented. Table 2. shows three different initial
perturbations used in the simulations; i.e. :
\begin{center}
\begin{tabular}{|l|}
\hline
$  n(x) = 1 + 0.01 \sin (k_0 x)$ \cr
\hline
$  n(x) = 1 + 0.01(-x+x^3)e^{-x^2} $ \cr
\hline
$ n(x)= 1 + 0.01\left\{
\begin{tabular}{rr}
$-1+x$ & $-1\le x < 0$ \cr
$1-x$ &   $0\le x \le 1$ \cr
$ 0 $ & Else where
\end{tabular}
\right.$
\cr
\hline
\end{tabular}
\end{center}
The reason for these choices is that the first perturbation
maintain harmonics with wave numbers very close to $k_{0}$ while
the latter two maintain harmonics more uniformly distributed in
the $k$ space. The most important reason for such choices was to
determine the impact of the initial perturbations on the final
wave spectra; strictly speaking the latter two are expected to
give rise to more uniform spectra. The initial velocity profiles
corresponding to these three profiles are drawn in Figs. 1(a), (b)
and (c). These velocity profiles indicate broader and more uniform
distribution of bulk flow velocities in the latter two; i.e. the
volume of phase space available to wave propagations are
considerably larger.
\begin{figure}
  %\vspace{1cm}
  \epsfxsize=9truecm
\centerline{\epsfbox{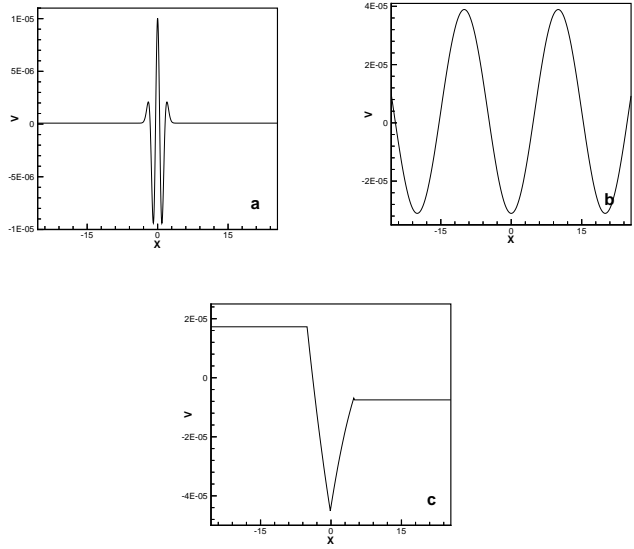}}
  \caption{Velocity profile for a) $n(x) = 1 + 0.01(-x+x^3)e^{-x^2}$,
  b)$n(x) = 1 + 0.01 \sin (k_0 x)$ and c) Saw-tooth function}
\end{figure}
Given these two facts though, the plots of the power
spectra\footnote{ The power spectrum is determined in two steps:
First, the spatial FFT is used in a quantity (e.g. E(x,t)) and
stored E($k_i$,t), next for each $k_i$ temporal FFT is performed
on E($k_i$,t)} of the modes versus $\omega$ (their frequency)
indicate very close agreement in all the cases; i.e. regardless of
the initially excited modes and phase velocities, most the allowed
$k$-space tends to get excited. This supports our earlier claim
that the discretization procedure and the numerical dispersion and
dissipation have in effect broadened and stabilized the initial
spectrum.

Finally the plots of the dispersion relation for a system size of
1024$\Delta$ with $\Delta=0.01$ are shown in Fig 2 and 3. The
close agreement between the analytic theory (solid lines) and the
model (circles) for wave numbers $k$ as large as 6 indicate
resolution of the modes with wave lengths of the order of grid
spacing with negligible numerical dispersion. Comparison of these
with the corresponding PIC simulations for a system 256$\Delta$
length (Fig. 4) clearly indicate resolution of much shorter
wavelengths here and considerably less numerical dispersion. This
is understandable since in the PIC models the finite particle size
effects introduce additional numerical dispersion which cause
smaller allowed $k$'s.
\begin{figure}
  %\vspace{1cm}
  \epsfxsize=9truecm
\centerline{\epsfbox{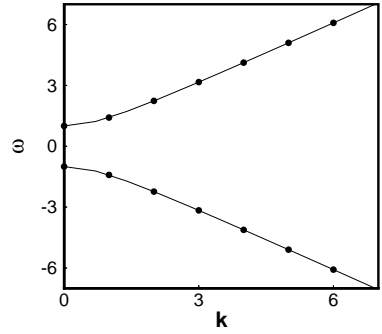}}
  \caption{Dispersion relation for Langmuir wave.}
\end{figure}
\begin{figure}
  %\vspace{1cm}
  \epsfxsize=6truecm
\centerline{\epsfbox{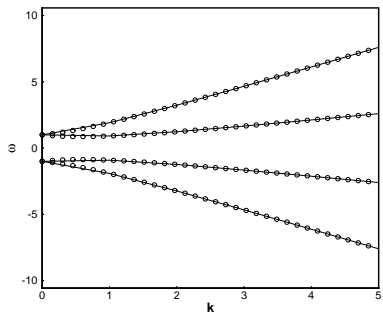}}
  \caption{Dispersion relation for Langmuir wave with Doppler
  effect}
\end{figure}
\begin{figure}
  %\vspace{1cm}
  \epsfxsize=6truecm
\centerline{\epsfbox{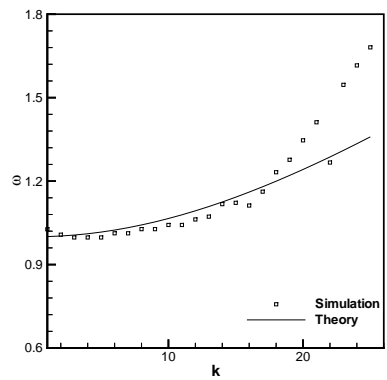}}
  \caption{Dispersion relation for Langmuir wave for a typical PIC
  simulation}
\end{figure}
One last remark about the cases corresponding to Figs. 2 and 3 is
that the latter involves the case in which the bulk plasma had an
initial flow velocity. Fig. 3 not only shows that the doppler
shifted waves also obey their respected dispersion relation, it
also shows how any "resulting" plasma flow could impact those
waves. That is if any nonzero average flow should arise from the
initial perturbations (i.e. if the scheme does not preserve
momentum conservation ) the dispersion relation would be impacted
as in Fig. 3. A glance at Fig. 2 though points that there could
not have been any doppler shift and therefore no net plasma flow
must have resulted from the initial perturbations. Calculations
also showed that $\langle v_f\rangle =0$ initially remained so to
round off errors throughout the simulation. So these plots also
probe the momentum to be conserved in the model.

\subsection{Wave Launching on the Boundary}

In the next example a wave is launched from the boundary and its
behavior is followed. Theoretically, recall that in an
unmagnetized plasma and in the linear regime the plasma shields
any incoming AC density perturbation whose frequency is less than
plasma frequency ($\omega_p$). This effect is shown in Fig. 5(b)
and Fig. 6. In this example the frequency of the applied density
perturbation is half of the plasma frequency. The wave is launched
at $x=-25 \lambda_D$. The amplitude of the density perturbation
has the following range: nonlinear (0.2,1.8) Fig. 5(a) and linear
(0.99,1.01) Fig. 5(b)\footnote{In these particular shots the wave
trough fall at launch points.}. The penetration depth is from
$x=(-25,-20)$ in the linear and $x=(-25,-15)$ in the nonlinear
case. Furthermore as Fig. 6(a) indicates, upon penetration, after
one wave period following the first crest ($x=-18$), the second
crest steepens with its wavelength decreasing to grid cell
scale.\footnote{The oscillations are numerical in nature. The
model should be modified to include FCT filter\cite{Boris} to
eliminate these spurious oscillations.} In the linear regime
though [Fig. 6(b)] no steepening can be seen.

\begin{figure}
  \vspace{2cm}
  \epsfxsize=10truecm
  \centerline{\epsfbox{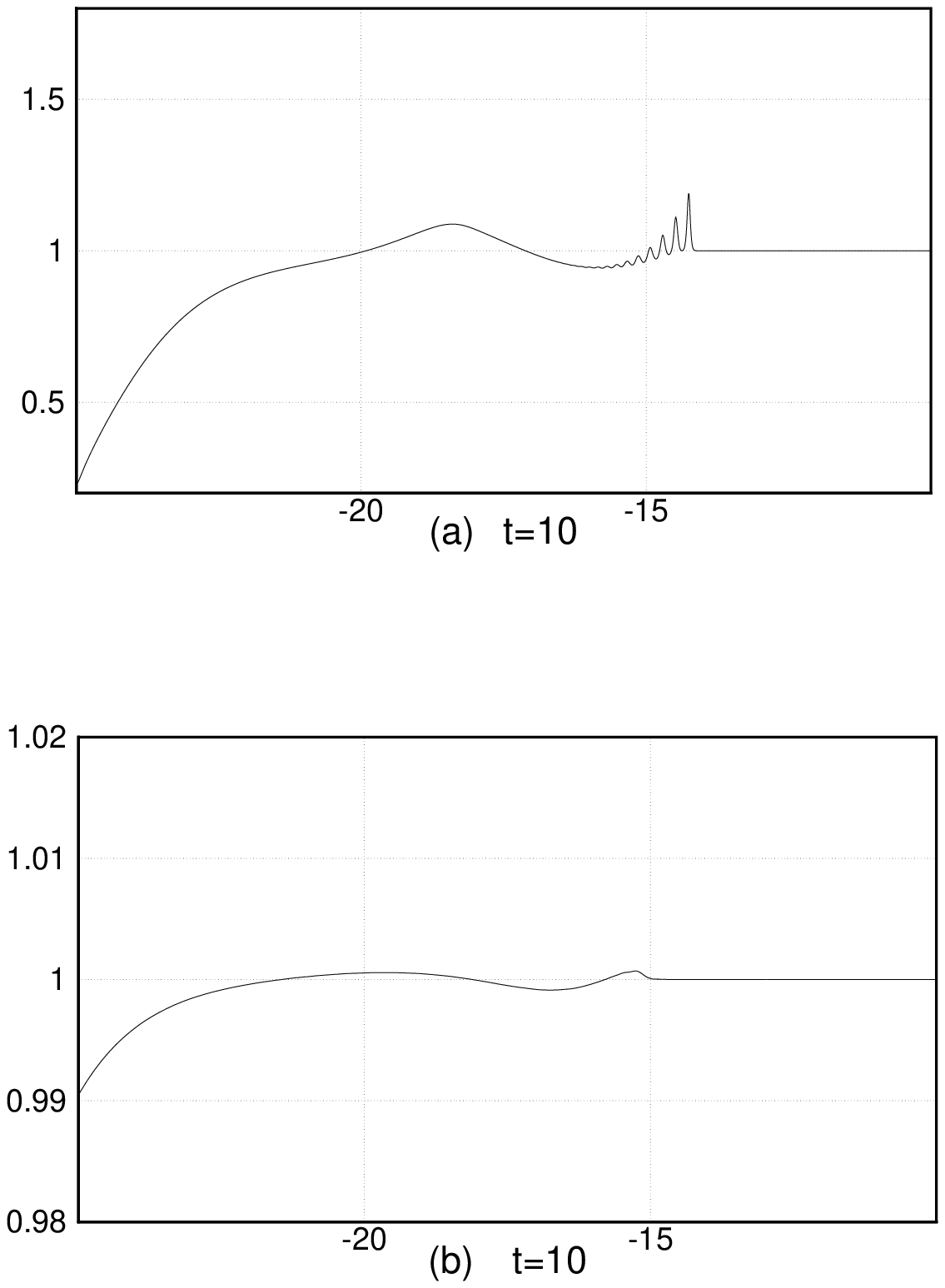}}
  \caption{Non-linear and linear penetration of electric field
  (both plots are sketched at t=10).
  a) Nonlinear case b) linear case  }
\end{figure}

In the other case, with the same initial condition (respect to
linear case), we launched a wave whose frequency was larger than
the plasma frequency( $\omega
> \omega_p$). This time the density perturbation propagated into the
plasma with its wavelength and amplitude unchanged as it
penetrated the plasma. Its behavior also conformed with the
analytic dispersion relation. The results are shown in Fig. 7.

\begin{figure}
  \vspace{3cm}
  \epsfxsize=12truecm
  \centerline{\epsfbox{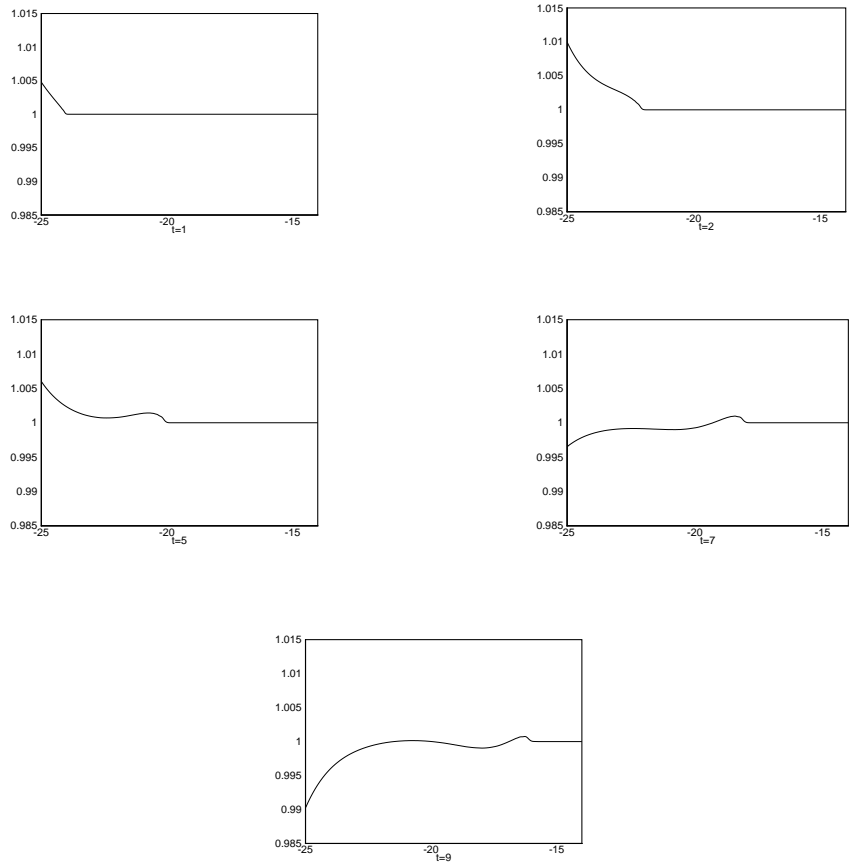}}
  \caption{Density versus the position when the external frequency is half of
  the plasma frequency. To give a time evolution feeling, they are plotted for
  five different normalized time. }
\end{figure}

\begin{figure}
  \vspace{3cm}
  \epsfxsize=12truecm
\centerline{\epsfbox{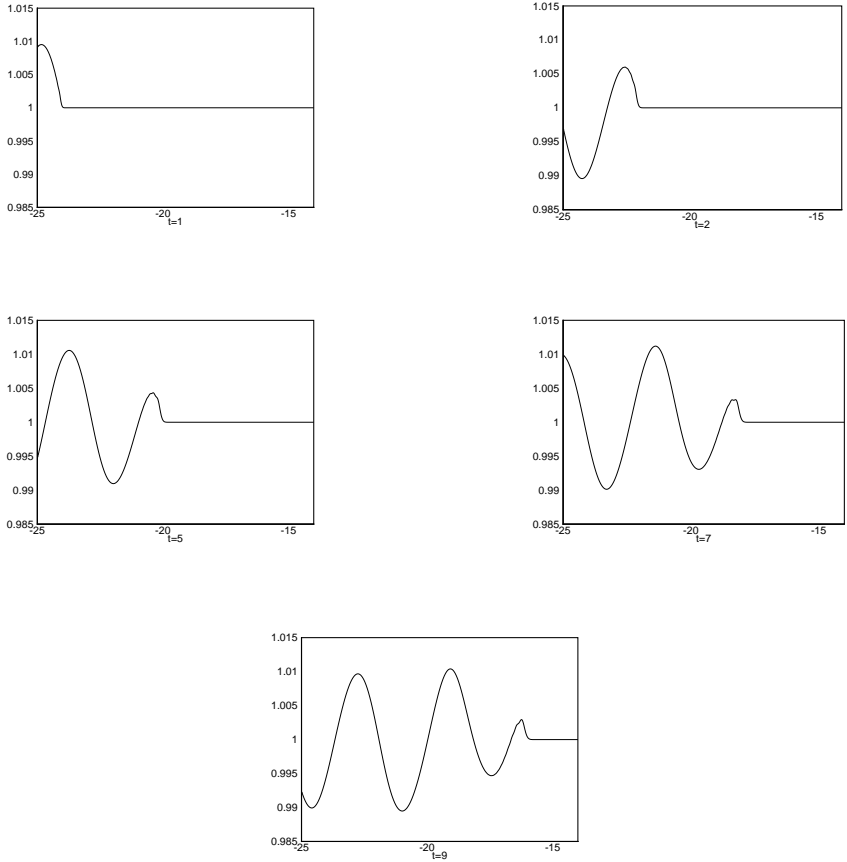}}
  \caption{Density versus the position when the external frequency is two times of
  the plasma frequency. To give a time evolution feeling, they are plotted for
  five different normalized time.}
\end{figure}

\subsection{Two Stream Instability}

As a more severe test of the code, we treated the two stream
instability. Although the instability arises under a wide range of
beam conditions, we shall consider only the simple case of two
countrastreaming uniform beams of electrons with the same number
density $n_0$. The first beam travels in the x direction with
drift velocity $v_d$ and the second beam in the opposite direction
with same drift velocity, i.e. the countrastreaming beams have the
same speed. The dispersion relation is as follows:
\begin{equation}
\frac{\omega_p^2}{(kv_d-\omega)^2}+\frac{\omega_p^2}{(kv_d+\omega)^2}=1
\end{equation}
where $\omega_p^2=4\pi e^2 n_0/m$ is the same plasma frequency for
both beams. One can then obtain the following expression for
$\omega^2$:
\begin{equation}
\omega^2=\omega_p^2+k^2v_d^2\pm
\omega_p{(\omega_p^2+4k^2v_d^2)}^{1/2}.
\end{equation}
This relationship between $\omega^2$ and $k^2$ is shown
graphically in Fig. 8. It is clear that, there exists a critical
wave number $k_c$ which separates the stable and unstable modes.
In fact , for $k^2<k_c^2$ two values of $\omega$ are complex, one
of which represents a growing wave; i.e. an instability. Moreover,
there exists a wave number $k_m$ that corresponds to the most
unstable mode.
\begin{figure}
  %\vspace{1cm}
  \epsfxsize=6truecm
\centerline{\epsfbox{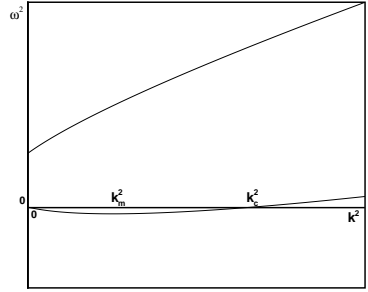}}
  \caption{Representation of relationship between $\omega^2$ and $k^2$.}
\end{figure}
These effects  are examined by the fluid code. In this case the
code was generalized to a two countrastreaming fluid model. As the
two countrastreaming beams emerging from the opposite ends meet
half way into the simulation box, a growing wavelike disturbance
develops. Figs. 9 and 10 show the evolution of this disturbance
for the cases with and without the pressure terms respectively. In
both cases the disturbance grows locally while in the latter it
also begins to propagate in both directions; i.e. a result of the
dispersion due to the pressure term.
\begin{figure}
  %\vspace{1cm}
  \epsfxsize=10truecm
\centerline{\epsfbox{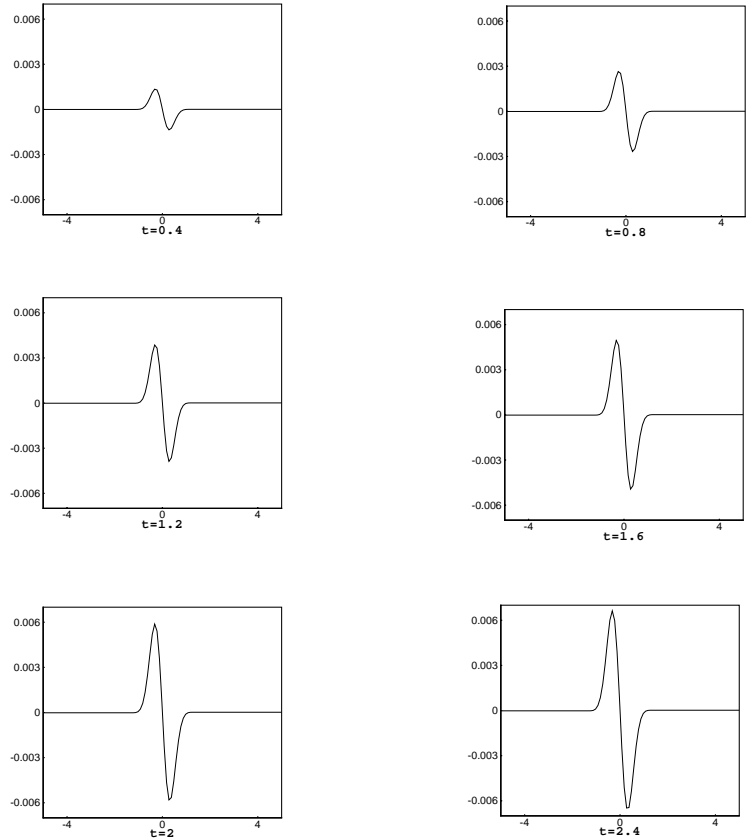}}
  \caption{Electric field versus the position in absence of pressure.
   Time is normalized by $\omega_p$.}
\end{figure}
\begin{figure}
  %\vspace{1cm}
  \epsfxsize=12truecm
\centerline{\epsfbox{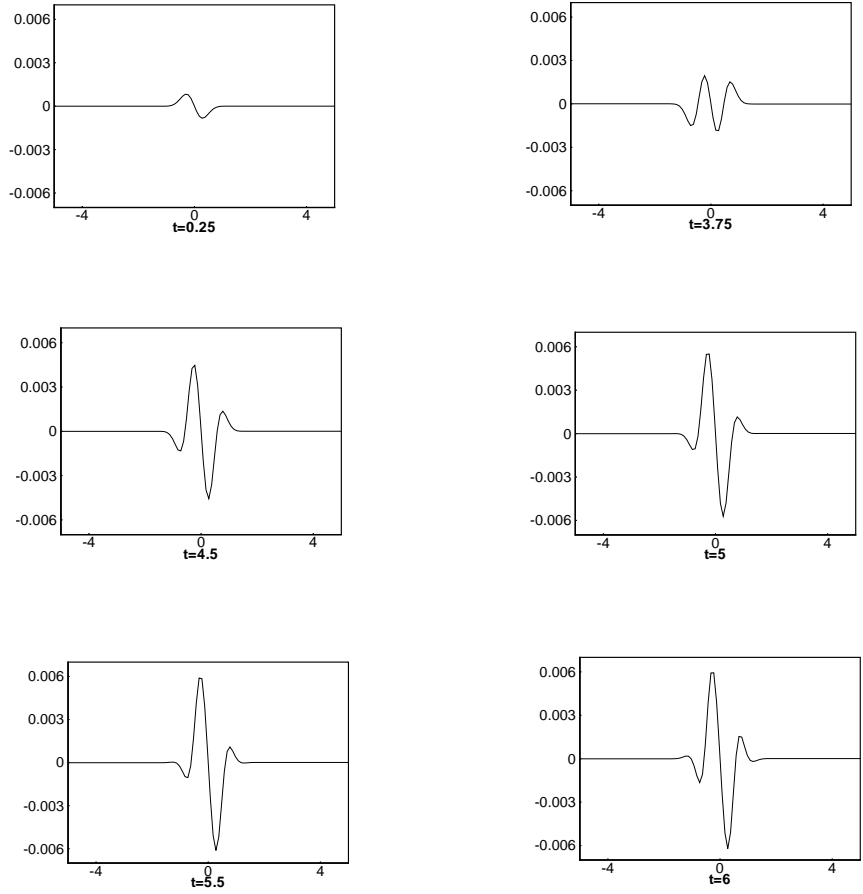}}
  \caption{Electric field versus the position in presence of pressure.
   Time is normalized by $\omega_p$.}
\end{figure}
Furthermore, the instability of each mode was investigated using
the mode energy discussed in the previous section: i.e.
\begin{equation}
P(k,t)={|E(k,t)|}^2
\end{equation}
The time derivative of this function with respect to $k$ is shown
in Fig. 11. As expected, there exists a critical wave number
bellow which unstable modes can grow. Furthermore we observed the
the most unstable mode corresponding to $k=k_m$ as the maximum in
the Fig. 11. Also the dynamic evolution of the beam-beam
interaction was observed as a movie and both the disturbance
growth and upstream propagations (when pressure term was included)
were observed.
\begin{figure}
  %\vspace{1cm}
  \epsfxsize=8truecm
\centerline{\epsfbox{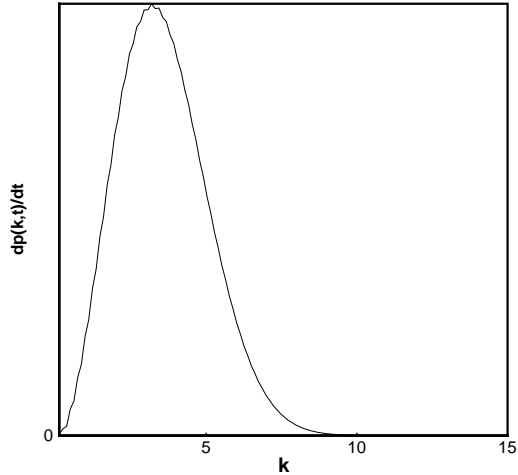}}
  \caption{$dp(k,t)/dt$ versus $k$. Cutoff and
  maximum wave numbers ($k_c$,$k_m$ ) are comparable with theory.}
\end{figure}

\section{conclusion}

The result of this paper demonstrates that fluid model can be used
to investigate any waves predicted by their basic set of equation.
This can include waves of kinetic nature with and without
dispersion with resolution far greater than the corresponding PIC
codes. It was demonstrated that appropriate initial perturbations
coupled with difference algorithms of sufficient but not excessive
numerical dispersion and dissipation can give rise to wave spectra
spanning all the allowed k-space. Many areas of plasma and or
space research can greatly benefit from these techniques.

\end{document}